\documentclass[%
 aip,
 amsmath,amssymb,
 reprint%
]{revtex4-1}
\usepackage[english]{babel}
\usepackage{graphicx}
\usepackage{psfrag}
\usepackage{amssymb}
\usepackage{amscd}
\usepackage{eucal}
\usepackage{color}
\usepackage{bm}
\usepackage{lipsum}
\usepackage{hyperref}
\usepackage{todonotes}
\usepackage{bm}
\newcommand*{\citen}{}
\DeclareRobustCommand*{\citen}[1]{%
  \begingroup
    \romannumeral-`\x 
    \setcitestyle{numbers}%
    \cite{#1}%
  \endgroup
}

\begin{document}

\newcommand{\J}{{\mathbb J}}
\newcommand{\Kf}{K\hspace{-0.8mm}f}

\title{Reconstructing Network Structures from Partial Measurements}
\author{Melvyn Tyloo}
  \altaffiliation[]{Corresponding author}
   \email{melvyn.tyloo@gmail.com}
\affiliation{Department of Quantum Matter Physics, University of Geneva, CH-1211 Geneva, Switzerland}
\affiliation{School of Engineering, University of Applied Sciences of Western Switzerland HES-SO, CH-1951 Sion, Switzerland.}
\author{Robin Delabays}
\affiliation{Automatic Control Laboratory, ETH Z\"urich, CH-8092 Z\"urich, Switzerland}
\affiliation{Center for Control, Dynamical Systems and Computation, UC Santa Barbara, Santa Barbara, CA 93106-5070 USA}
\author{Philippe Jacquod}
\affiliation{Department of Quantum Matter Physics, University of Geneva, CH-1211 Geneva, Switzerland}
\affiliation{School of Engineering, University of Applied Sciences of Western Switzerland HES-SO, CH-1951 Sion, Switzerland.}

\date{\today}

\begin{abstract}
The dynamics of systems of interacting agents is determined by the structure of their coupling network. The knowledge of the latter is therefore highly desirable, for instance to develop efficient control schemes, to accurately predict the dynamics or to better understand inter-agent processes. In many important and interesting situations, the network structure is not known, however, and previous investigations have shown how it may be inferred from complete measurement time series, on each and every agent. These methods implicitly presuppose that, even though the network is not known, all its nodes
are. Here we investigate the different problem of inferring network structures within the observed/measured agents. 
For symmetrically-coupled dynamical systems close to a stable equilibrium, we establish analytically and illustrate numerically that 
velocity signal correlators encode not only 
direct couplings, but also geodesic distances in the coupling network, within the subset of measurable agents.
When dynamical data are accessible for all agents, our method is furthermore algorithmically more efficient than the traditional ones, because it does not rely on matrix inversion. 
\end{abstract}
 
\maketitle
\begin{quotation}
Network inference is useful for complex systems as diverse as biological networks, power grids, information and social networks to name but a few. Time series of the degrees of freedom are typically obtained through experiments or monitoring on systems that are most of the time only partially accessible. Indeed, in some systems the number of interacting agents is too large so that it is impossible to monitor all of them, or some agents might be hidden or one only needs information on a specific part of a larger coupled system. In this work we propose a method to infer the network structure within a set of accessible agents, for which time series of the degrees of freedom are measurable. The observed system is subjected to noise that might come from environmental degrees of freedom that were neglected or from external perturbations. We analytically connect the two-point correlators of the velocity deviations to the underlying coupling network for a general class of symmetrically-coupled systems close to a stable equilibrium.
\end{quotation}

\section{Introduction}
Network science -- the field that studies complex, networked systems~\cite{New03} -- has seen an enormous growth of activity in recent years.
More and more diverse systems of physical, life and human sciences are analyzed through larger and larger models of agents connected to one another,~\cite{Bar16} thanks in large part to the ever-increasing capacity for data mining and processing.~\cite{Hil11} As a matter of fact, network science draws heavily on data science, however,
it also draws on analytical methods, most notably of statistical physics, graph theory and dynamical systems.

Approaches combining analytical and data-based 
approaches generally compensate for the weaknesses of one with the strengths of the other and are currently 
extensively applied to solve challenging problems of network science. 
One such challenging problem is to reconstruct the structure of a priori unknown 
networks from sets of dynamical measurement data of its agents. Time series recording the agents 
dynamics are used to infer the topology of their coupling network when the direct observation of the latter is
impossible.~\cite{Wan16b,Bru18} The gained knowledge of the coupling network is then used to evaluate
the state of the system more precisely, to predict its future evolution, to anticipate extreme behaviors,
to implement control schemes, to deduce inter-agent processes and so forth. The problem is of particular interest for
noisy social networks which change over short time scales,~\cite{Sek16}
interconnected power grids and information networks
whose topology is regularly modified by line faults and reroutings,~\cite{Mac08,Bac13,Kir16,Tyl19}
or gene regulatory networks made of such huge numbers of proteins and genes that the precise structure of their interaction network is inaccessible.~\cite{Bra03,Gar03,Mar10} In all these examples, 
it is particularly important to have inference methods that are resilient against missing data and that can still reliably provide 
partial network structures in the case of incomplete measurements, i.e. when not all agents are measurable. 


There is a rather vast literature on network inference from dynamical measurements of agents and many data-based methods 
have been constructed.
Early approaches use a probe injection signal and measure the response dynamics of the agents.~\cite{Yu06,Tim07,Yu08,Don13,Bas18,Fur19,Tyl21} 
The successful reconstruction of the network topology, through e.g., the Laplacian matrix, the Jacobian matrix of dynamical flows or the adjacency matrix, requires then  
not only to record the dynamics of all agents, but also that one can control and inject specially tailored probe signals. 
Less demanding passive methods have been devised, which rely only on observations of the agents dynamics. 
Some are based on the optimization of a likelihood~\cite{Hoa19} or cost function, and require a
computation time that scales at least as ${\cal O}(n^3)$,~\cite{Mak05}
or even as ${\cal O}(n^4)$,~\cite{Sha11,Pan19}
with the number $n$ of agents. 
To reconstruct large networks, a computationally more efficient method is therefore
highly desirable.
Lighter, probabilistic approaches identify likely couplings between pairs of agents from statistical properties of the corresponding pairs of trajectories.~\cite{Dah97,Sam99,New18,Pei19,Leg19,Ban19}
A different and rather efficient approach extracts the network topology from 
the $n (n-1)/2$ two-point correlators of pairs of agent trajectories in systems subjected to  
white~\cite{Ren10,Wan12,Chi17} or correlated noise.~\cite{Che162,Tam18} 
The method is in principle quite accurate, however it assumes that every agents in the system are measurable.
Because in many systems, measurements of only subsets of agents are possible, or because  
one often cannot be sure that all nodes are actually known and recorded, 
it is important to develop a reconstruction method that can 
extract even partial but 
reliable information on the network from dynamical data over a subset of the agents. 
In this manuscript we construct such a method 
for a general class of symmetrically-coupled systems close to a stable equilibrium. In contrast to purely data-based approaches such as machine learning algorithms or optimization techniques, our method connects statistical properties of time series data to the coupling between agents. Therefore it does not rely on any training set, nor on minimization solvers, but only on sufficiently long time series.

\section{Results}

We consider general dynamical systems defined by coupled ordinary differential equations, in the vicinity of a stable fixed point solution. 
Stability means that upon not too large deviations, the system remains close to the fixed point. 
Consider now that the system is initially there, 
but is subjected to some noisy perturbation. The latter may originate from simplifications in constructing the model, 
from the coupling to unavoidable environmental degrees of freedom or from a deliberately applied perturbation.~\cite{Kam76} Assuming that the noise is sufficiently weak, the system  
wanders stochastically about, but remains close to the fixed point for a long time.~\cite{Del17b,Tyl18c} We 
record the dynamics of the agents and, from these time series, compute two-point velocity correlators, $\langle \delta \dot{x}_i \delta \dot{x}_j \rangle$, between 
measurable agents $i$ and $j$. Our key observation is that, unlike the position correlators considered so far,~\cite{Ren10,Wan12,Chi17,Che162,Tam18} 
velocity correlators contain direct information on the network Jacobian matrix of dynamical flows  [see Eq.~(\ref{eq:dis2}) below].
The method therefore
enables the direct reconstruction of network structures, without any matrix inversion.
This apparently minor improvement impacts network inference very significantly and positively in that,
first, and most importantly, avoiding the matrix inversion enables to still recover partial information on the network matrix
 when only a subset of the agents is measurable;
second,  matrix inversion being a computationally costly operation, our method is scalable to larger networks; 
third, our method is able to identify not only direct couplings, but also the geodesic distance between pairs of not too distant observable agents.
Additionally, we show below that the method can efficiently determine 
topological changes in time-evolving networks. The price to pay for these improvements is moderate. As
a matter of fact, 
we show in the Supplementary Information that measuring velocity instead of position correlators does not
require a prohibitively fine time sampling of the dynamics, and that our method is more robust against measurement noise.

\begin{figure*}
 \centering
 \includegraphics[width=0.95\textwidth]{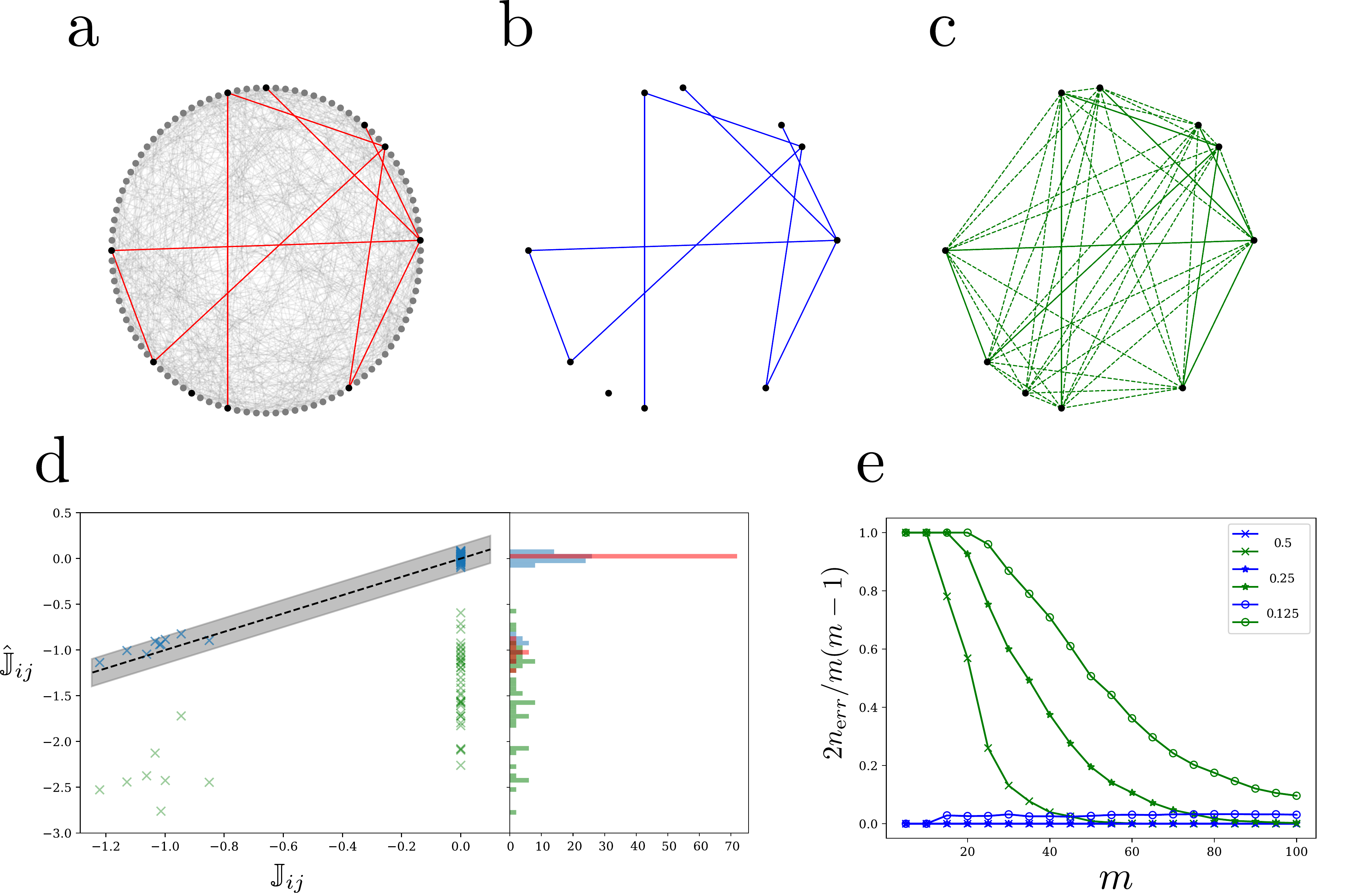}
 \caption{
 \textbf{Partial network inference from partial measurements.} (a) Erd\H os-R\'enyi random network~\cite{Bar16} with $n=100$ agents subjected to the dynamics defined by Eqs.~(\ref{eq:dyn0}), (\ref{eq:F}) and (\ref{eq:laplacian}) with $f(x)=\sin(x)$\, and nonzero matrix elements 
 $\J_{ij} \in [0.6,1.3]$. Only the $m=10$ black nodes are measurable. They are directly connected by the red edges. (b) Edges inferred using our method, Eq.~(\ref{eq:est1}). 
 The inferred partially reachable network (blue) is the same as the true, accessible one (red) in panel (a). (c) Edges inferred from the
 method of Ref.~\citen{Ren10} (see Supplementary Information) applied outside of its regime of validity to the subset of accessible nodes. Solid green edges are those correctly inferred while dashed green edges are 
 incorrectly predicted to exist. (d) Elements of the Jacobian matrix of dynamical flows between the $10$ black nodes in panel (a). All blue crosses are within the grey area corresponding to $\J_{ij}\pm 0.1$\,. Red histogram bars give the distribution of true matrix elements; blue crosses and histogram bars are results from Eq.~(\ref{eq:est1}); green crosses and histogram bars correspond to the method of Ref.~\citen{Ren10}. (e) Fraction of errors $2 n_{err}\big/m(m-1)$ of inferred off-diagonal elements 
 $\hat{\J}_{ij}$ as a function of the number of accessible nodes $m$\,. Blue symbols correspond to our method, Eq.~(\ref{eq:est1}); green symbols correspond to the method of Ref.~\citen{Ren10}. Different symbols correspond to different tolerances, $\epsilon_{\rm err}>|\hat{\J}_{ij}-\J_{ij}|$\,, for correctly inferred couplings.}
 \label{fig:fig1}
\end{figure*}

\begin{figure*}
 \centering
 \includegraphics[width=0.95\textwidth]{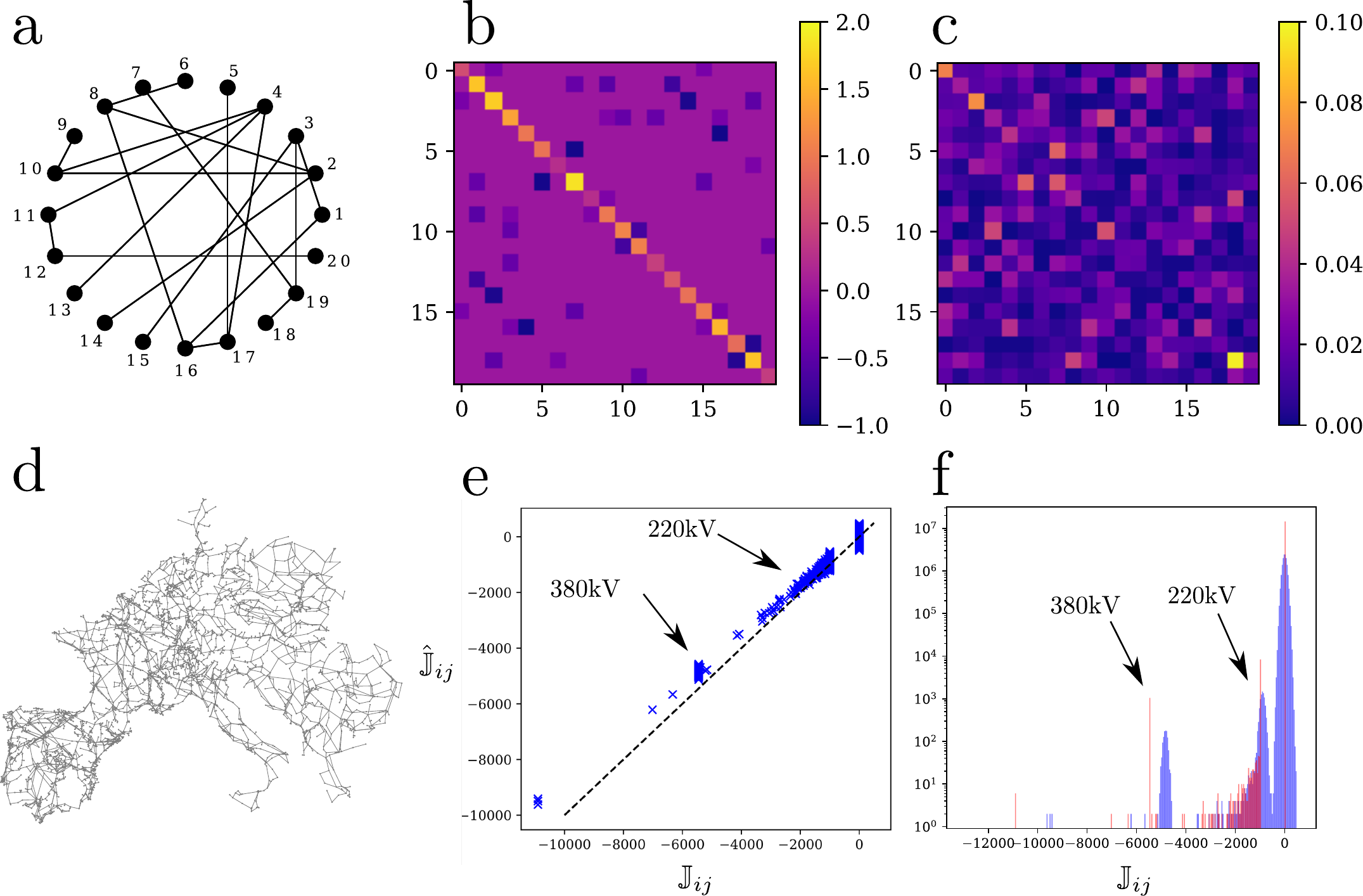}
 \caption{\textbf{Full network reconstruction from complete sets of measurement.} (a) Complex network used to generate time series following the dynamics defined by Eqs.~(\ref{eq:dyn0}) and (\ref{eq:F}) with $f(x)=\sin(x)$\,, in the limit $\lambda_\alpha \tau_0\ll 1$ of short noise correlation time. (b) Color-coded true Jacobian matrix of the dynamical flows for the network in panel (a). (c) Color-coded
difference $\Delta \J_{ij}=|\hat{\J}_{ij}-\J_{ij}|$ between matrix elements of the inferred ($\hat{\J}$)
and true ($\J$)
Jacobian matrices. Note the difference in colorscale between panels (b) and (c).
(d) PanTaGruEl model of the synchronous power grid of continental Europe~\cite{Tyl19} used to generate time series following the dynamics defined by Eqs.~(\ref{eq:dyn0}) and (\ref{eq:F}) with $f(x)=\sin(x)$\,, in the limit $\lambda_\alpha \tau_0\ll 1$ of short noise correlation time. The complete network has $n=3809$ nodes and 4944 edges. (e) Complete inference  for the PanTaGruEl network with all agents accessible to measurement. Blue crosses plot the inferred matrix elements $\hat{\J}_{ij}$ from Eq.~(\ref{eq:est1}), against their real value $\J_{ij}$\,. (f) Histogram of the inferred velocity (blue) and of the true Jacobian matrix (red) of dynamical flows. The method satisfactorily infers the Jacobian matrix elements that vary over more than one order of magnitude, and furthermore identifes the two types of edges, corresponding to different voltage levels of 220 and 380 kV [arrows in panels (e) and (f)]. The separation between low-valued inferences (corresponding to non-existant edges) 
 and higher-valued ones is obvious. The small but still 
 significant inference imprecision is due to computational limits for generating velocity time series by simulating the dynamics
 of this large network, and not to our inference method. 
}
 \label{fig:fig2}
\end{figure*}

The power of our method to infer partially accessible network structures is illustrated in Fig.~\ref{fig:fig1} for a dynamical system of $n=100$ agents on a random Erd\H os-R\'enyi network. 
Existing network couplings between pairs of $m=10$ measurable agents are shown in red in panel (a). 
An observer, unaware that they have access to a fraction of the network agents only and
who would apply the position correlator method of Refs.~\citen{Ren10,Wan12,Chi17,Che162,Tam18}
outside its range of validity, would generally conclude that 
all pairs of agents are directly coupled, because the method relies on a matrix inversion (see Supplementary Information).
 This is shown in panel (c). 
Furthermore, the position correlator method also fails quantitatively in that it predicts coupling strengths that are too large by a factor of up to three. 
These shortcomings do not affect our method, however, which correctly predicts qualitatively and quantitatively the couplings between the $m=10$ 
measurable agents [blue lines and histogram in panels (b) and (d)]. When all agents are accessible,
our method finally captures the full network structure with high precision. This is illustrated in Fig.~\ref{fig:fig2}. 

\subsection{Network-coupled dynamical systems.} 
We consider a system of $n$ agents whose coordinates are cast in a vector ${\bf x} \in\mathbb{R}^n$. Their dynamics is governed by a generic 
autonomous ordinary differential equation
\begin{equation}\label{eq:dyn0}
\dot{\bf x}(t) = {\bf F}[{\bf x}(t)] \, ,
\end{equation}
Assume next that this equation has a stable fixed point solution ${\bf x}^*$, i.e. ${\bf F}[{\bf x}^*]=0$, that ${\bf F}=(F_1, \ldots F_n)$ is a real vector function that is 
differentiable about ${\bf x}^*$, and that the
Jacobian matrix of the dynamical flows, $\J_{ij}(\bm{x}^*)=-\partial F_i(\bm{x}^*)/\partial x_j$, is real symmetric and positive semidefinite at ${\bf x}^*$. The pairwise couplings between agents is encoded in $\J_{ij}$ and the 
latter condition guarantees the
stability of the fixed point solution under not too strong perturbations. Assume finally that the system is subjected to a noisy perturbation ${\bm \xi}(t)$,
starting initially at the fixed point. When the noise perturbation is sufficiently weak, the system remains in the vicinity of the fixed point for long times~\cite{Del17b,Tyl18c} and
its dynamics is well captured by the linearized ordinary differential equation
\begin{equation}\label{eq:dyn_lin}
 \bm{\delta} \dot{\bm x} = -\J(\bm{x}^*)\,\bm{\delta x} + \bm{\xi} \, ,
\end{equation}
governing the vector of deviation coordinates $\bm{\delta x}=\bm{x}-\bm{x}^*$.
Despite its simplicity and the assumptions on which it is based, Eq.~(\ref{eq:dyn_lin})
is used to analyze a wide variety of systems, such as 
electric power grids subjected to fluctuations of loads~\cite{Bac13,Tyl18a,Ron18},
consensus algorithms in computer science,~\cite{Lyn97} opinion dynamics  
in social sciences,~\cite{Heg02} vehicle platoon formation and stability in trafic modeling and control~\cite{Bam12} or contagion dynamics at early stages of epidemia.~\cite{Col07,Bro13}


The matrix elements $\J_{ij}({\bm x}^*)$ contain the information we want to extract on the coupling network between agents $i$ and $j$. It is the matrix we want to reconstruct. 
Under our assumptions that it is real symmetric and that ${\bm x}^*$ is a stable fixed point, $\J({\bm x}^*)$ has real, nonnegative eigenvalues, $0\le \lambda_1 \le \lambda_2\leq ...\leq \lambda_n$, associated to a complete orthonormal basis of eigenmodes $\{ \bm{u}_\alpha \}_{\alpha=1}^n$.
Eq.~(\ref{eq:dyn_lin}) is solved by a spectral expansion of the displacements over this basis, $ \bm{\delta x}(t) = \sum_\alpha c_\alpha(t)\bm{u}_\alpha\,$. This
leads to a set of uncoupled Langevin equations with solution~\cite{Tyl18a}
\begin{equation}\label{eq:sols}
 c_\alpha(t) = e^{-\lambda_\alpha t}\int_0^te^{\lambda_\alpha t'} \bm{\xi}(t')\cdot \bm{u}_\alpha{\rm d}t'\, 
\end{equation} 
for the coefficients of the spectral expansion.
To calculate the equal time, two-point velocity correlator $ \langle\delta\dot{x}_i(t) \delta\dot{x}_j (t)\rangle$ between agents $i$
and $j$ we next need to specify the noise distribution.

\subsection{Network reconstruction from ambient noise.}
The perturbation noise in Eq.~(\ref{eq:dyn_lin}) is unavoidable. This is in particular so since 
real systems are often too complicated to be exactly modeled by a tractable 
function ${\bf F}[{\bf x}]$. Therefore, noise is often introduced to mimic the effect of neglected terms or to model uncapturable environmental degrees of freedom~\cite{Kam76}. It is reasonable to 
assume that this ambient noise has fast decaying spatial correlations and is characterized by some finite correlation time $\tau_0$. 
Accordingly, we model ${\bm \xi}(t)$ by an Ornstein-Uhlenbeck process defined by its first two moments 
\begin{eqnarray}\label{eq:noise}
 \langle\xi_i(t)\rangle &=& 0\, ,\nonumber  \\ 
 \langle\xi_i(t+\Delta t/2)\xi_j(t-\Delta t/2)\rangle &=& \xi_0^2 \, \delta_{ij} \exp\left(-|\Delta t|/\tau_0\right)\, , \qquad
\end{eqnarray} 
where $\xi_0$ is the noise standard deviation and the brackets denote ensemble averaging over noise realizations or a large enough observation time, $\langle ... \rangle = {\rm lim}_{T \rightarrow \infty} T^{-1}\int_0^T ... \, {\rm d}t$.
Ambient noise originates from couplings to an environment that is large by definition. Accordingly, one standardly assumes that $\tau_0$ is one
of, if not the shortest time scale in the problem. In our discussion we take $\tau_0$ as an independent parameter, but we often assume below that 
the physically relevant limit is $\tau_0 \rightarrow 0$. 

In this paper, we specialize to sets of agents with two-body interactions, $F_i = \omega_i + \sum_{j \ne i} {\cal F}_{ij}$, whose coupling network can be inferred from 
the linearized dynamics close to the fixed point solution ${\bm x}^*$ of Eq.~(\ref{eq:dyn0}). Accordingly, only the first two moments of the noise need to 
be specified to infer the coupling network through ${\cal F}_{ij}$. We note that our method 
can be extended to coupling networks with higher order interactions between three or more agents, in which case
one needs however to specify higher order moments of ${\bm \xi}(t)$. Also worth noticing is that Eq.~\eqref{eq:dyn_lin} 
satisfies conditions for dynamical structure reconstruction.~\cite{Gon08}

Earlier approaches reconstruct first the pseudo-inverse Jacobian $\J^\dagger$ from 
two-point position correlators $\langle\delta x_i(t) \delta x_j (t)\rangle$ derived from
dynamical measurements over all agents.~\cite{Ren10,Wan12,Chi17,Che162,Tam18}
Here, we consider instead two-point {\it velocity} correlators, $\langle\delta\dot{x}_i(t) \delta\dot{x}_j (t)\rangle$, which
enables us to directly reconstruct $\J$, without any matrix inversion, as we will show shortly.
We consider the long-time influence of the noise perturbation, after all initial transient behaviors have relaxed. 
Expanding the velocities over the eigenmodes of $\J$, $ \bm{\delta \dot x}(t) = \sum_\alpha \dot c_\alpha(t)\bm{u}_\alpha\,$, and 
using Eqs.~(\ref{eq:sols}) and (\ref{eq:noise}), it is straightforward to obtain (See Supplementary Information)
\begin{equation}\label{eq:cor}
{\rm lim}_{t \rightarrow \infty}  \langle\delta\dot{x}_i(t) \delta\dot{x}_j (t)\rangle = \xi_0^2 \Big( \delta_{ij} - \sum_{\alpha} u_{\alpha,i}u_{\alpha,j}\frac{\lambda_\alpha\tau_0}{1+\lambda_\alpha\tau_0} \Big) ,
\end{equation}
where $u_{\alpha,i}$ is the $i^{\rm th}$ component of the $\alpha^{\rm th}$ eigenmode. 

Eq.~(\ref{eq:cor}) connects the long-time velocity correlator to the eigenmodes and eigenvalues of the Jacobian matrix $\J$. To extract network structures
from it, we recall that the matrix elements of the $k^{\rm th}$ power of $\J$ read $(\J^k)_{ij}=\sum_\alpha \lambda_\alpha^k \, u_{\alpha,i} u_{\alpha,j}$.  
Taylor-expanding Eq.~(\ref{eq:cor}) in the limit of short correlation time, $\lambda_\alpha\tau_0 < 1$, then gives
\begin{equation}\label{eq:dis2}
{\rm lim}_{t \rightarrow \infty}  \langle\delta\dot{x}_i(t) \delta\dot{x}_j (t)\rangle = \xi_0^2 \left[ \delta_{ij} + \sum_{k=1}^\infty (-\tau_0)^k(\J^k)_{ij} \right] \, .
\end{equation}
In the opposite limit of $\lambda_\alpha\tau_0 > 1$, another Taylor-expansion connects the velocity correlator to powers of the inverse Jacobian
instead (See Supplementary Information).
As argued above, if the noise perturbation arises from a large, fast-varying environment, the limit $\lambda_\alpha\tau_0 < 1$ of short noise correlation time is 
expected to be physically more relevant. Accordingly, we base our network reconstruction approach on Eq.~(\ref{eq:dis2}).

\subsection{Direct network reconstruction.} 
In the limit of very short noise correlation time, $\lambda_\alpha\tau_0\to 0$, only the $k=1$ term in Eq.~(\ref{eq:dis2}) matters, which 
gives
\begin{equation}\label{eq:est1}
 \hat{\J}_{ij} = (\delta_{ij} - \langle\delta\dot{x}_i\delta\dot{x}_j\rangle/\xi_0^2) \, \tau_0^{-1}\, .
\end{equation}
The Jacobian matrix $\J$ of dynamical flows is directly given by the long-time velocity correlator. Eq.~(\ref{eq:est1}) enables the complete
reconstruction of the network when all nodes are measurable. It is important to realize that this is done passively, i.e. solely by measuring the 
dynamics of the agents. In particular, the method does not require to control the perturbation. For full network reconstruction, Eq.~(\ref{eq:est1}) 
improves on earlier approaches in that the considered velocity correlators directly give the matrix elements of $\J$ and not of its inverse. This is algorithmically advantageous, especially in systems with many agents. If one is only interested in the structure of the Jacobian matrix then $\tau_0$ and $\xi_0$ do not need to be known. If one wants a more quantitative inference, $\tau_0$ can be extracted from the frequency spectrum of the time series, while the noise amplitude is obtainable from 
the variance of the agent coordinate at a single node, $\langle\delta\dot{x}_i^2\rangle=\xi_0^2$\,.
For full reconstruction, the method is
numerically illustrated in Fig.~\ref{fig:fig2}. Quite remarkably, one sees in Fig.~\ref{fig:fig2}(e-f) that our method correctly identifies different magnitudes of the Jacobian matrix elements.

Eq.~(\ref{eq:est1}) furthermore enables to identify direct connections between nodes without the need to reconstruct the full matrix. 
This is especially important when one has access to only a subset of the agents in the network or if one wants to check the connectivity only within a particular subset of the nodes. 
Then, our approach allows us to 
infer all direct connections between pairs of agents within those subsets. 
This is illustrated in Fig.~\ref{fig:fig1} where one sees that Eq.~(\ref{eq:est1}) accurately reconstructs the direct couplings, including their magnitude,
between the $m=10$ measurable agents in an Erd\"os-R\'enyi network of $n=100$ agents. An observer who would not know that 
only a subset of the agents is being recorded and who would apply approaches based on 
position correlators outside their range of applicability,
would wrongly conclude that the coupling between these $m=10$ agents is all-to-all. This is so, since these methods first construct
the inverse of the network matrix, and then invert that partially known matrix.
\begin{figure*}
 \centering
 \includegraphics[width=0.95\textwidth]{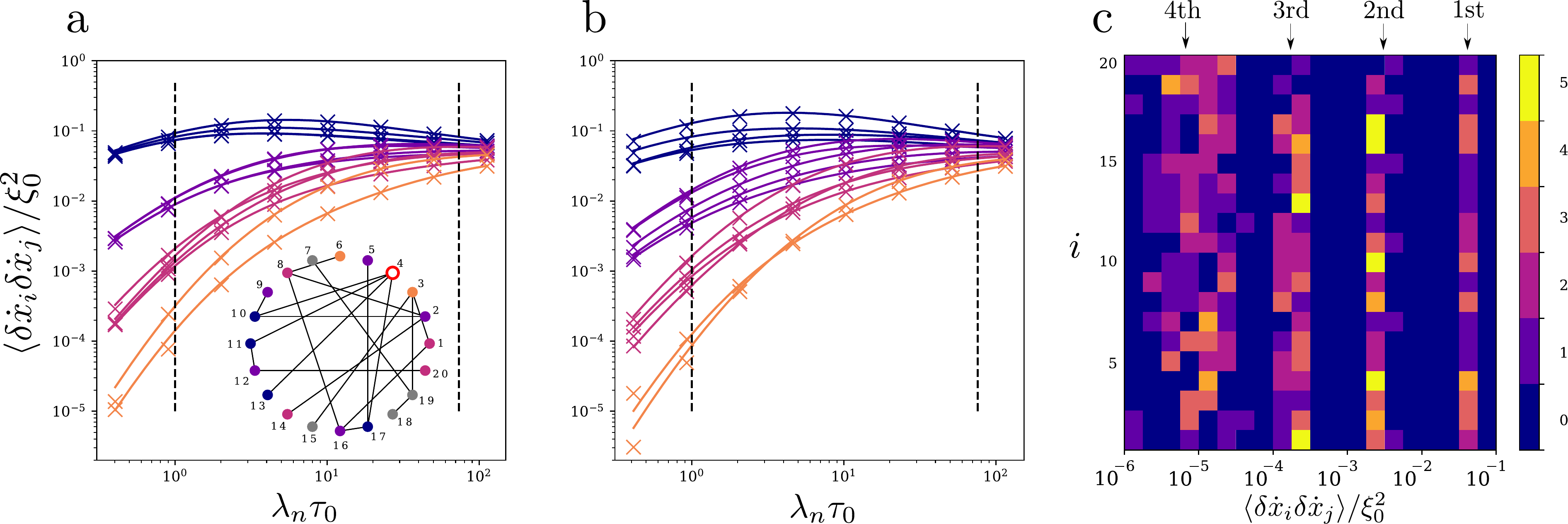}
 \caption{\textbf{Inference of geodesic distances.} Long-time velocity correlator of Eq.~(\ref{eq:cor}) between agent $\#4$ [open red circle in the network
 map in the inset of panel (a)] and all 
 other agents up to its $4$th neighbors. The geodesic distance between pairs of considered agents is color-coded. 
 (a) Network with uniform couplings. (b) Network with inhomogeneous couplings with strengths varying by a factor of 4.5. 
 Vertical dashed lines indicate the boundaries $\tau_0=\lambda_n^{-1}$ (left) and $\tau_0=\lambda_2^{-1}$ (right) from the spectrum of the Jacobian. 
 (c) Complete reconstruction of the 1st to 4th neighbors from Eq.~(\ref{eq:cor}) for the network in the inset to panel (a) with homogeneous
 couplings.
For each agent $i=1, \ldots 20$, and each correlator value,  
 the color plot gives the number of agents $j\neq i$ with that correlator value for $\lambda_n \tau_0=0.4$ [smallest value in panel (a)].
} 
\label{fig:fig3}
\end{figure*}

\subsection{Inferring geodesic distances.} 
When the noise correlation time is small, but finite, one can extract further important information from
Eq.~(\ref{eq:est1}), beyond direct network couplings. Suppose that two measurable agents $i \ne j$ are located a geodesic distance $q$
away from each other. This means that the shortest path $i \rightarrow k_1 \rightarrow k_2 \rightarrow \ldots k_{q-1} \rightarrow j$
from $i$ to $j$ goes through $q$ direct network couplings. Accordingly, $q$ is the lowest exponent for which $(\J^{q})_{ij} \ne 0$ and
therefore, for such pairs $i \ne j$ one has, instead of (\ref{eq:dis2}),
\begin{equation}\label{eq:geo}
 \langle \delta\dot{x}_i\delta\dot{x}_j \rangle = \xi_0^2 \sum_{k=q}^\infty(-\tau_0)^k\left(\J^k\right)_{ij} \, .
\end{equation}
This makes it possible to determine the geodesic distance $q$ between any measurable pair of nodes $(i,j)$ as long as 
\begin{equation}\label{eq:ineq}
{\rm min}_{l,m} (\J^{q-1})_{lm} \tau_0^{-1} \gg (\J^q)_{ij} \gg {\rm max}_{l,m} (\J^{q+1})_{lm} \tau_0 \, ,
\end{equation} 
where the minimum (resp. maximum) is taken over pairs $(l,m)$ of nodes with geodesic distance $\leq q-1$ (resp. $\geq q+1$). 
When Eq.~(\ref{eq:ineq}) holds, pairs of nodes with geodesic distance $q$ have noise correlators sufficiently away from those with geodesic 
distances $q-1$ and $q+1$ that one can identify them. 

Inference of geodesic distances between pairs of measurable agents is illustrated in Fig.~\ref{fig:fig3} for a small random network with $n=20$ agents. When the 
noise correlation time $\tau_0$ is sufficiently small, one sees that the values of long-time 
velocity correlators coalesce into distinct clusters. Each cluster corresponds to agents located
a fixed geodesic distance away from the chosen agent. Cluster correlator values decrease with increasing geodesic distance, which allows to 
infer the latter. Remarkably, the method works even when the network has nonhomogeneous couplings
[see panel (b)], and is limited only by correlator values becoming
smaller and smaller as the geodesic distance increases. 

We found that the inference of geodesic distances is in practice limited to identifying the first few neighbors.
This is so because, first, it requires short correlation times and second, 
from Eq.~(\ref{eq:dis2}), pairs of $k^{\rm th}$ neighbor agents $(i,j)$ have a noise correlator given by
\begin{equation}\label{eq:disk}
 \langle\delta\dot{x}_i\delta\dot{x}_j\rangle = \xi_0^2 (-\tau_0)^k(\J^k)_{ij} +{\cal O}[\tau_0^{k+1} (\J^{k+1})_{ij}] \, .
\end{equation}
As $k$ increases, the correlator therefore becomes smaller and smaller, until it eventually is smaller than its statistical standard deviation, at which 
point geodesic distances can no longer be inferred.
For the networks we investigated [see Fig.~\ref{fig:fig3}]
we have found that geodesic distances up to $k=3, 4$ can typically be inferred. 
As a remark concluding this paragraph, 
we stress that Eqs.~(\ref{eq:geo}) and (\ref{eq:ineq}) allow to extract geodesic distances
between pairs of measurable agents, even when only a subset of the agents is accessible. In that situation, 
powers of the partially inferred Jacobian would systematically overestimate geodesic distances.

\begin{figure*}
 \centering
 \includegraphics[width=1\textwidth]{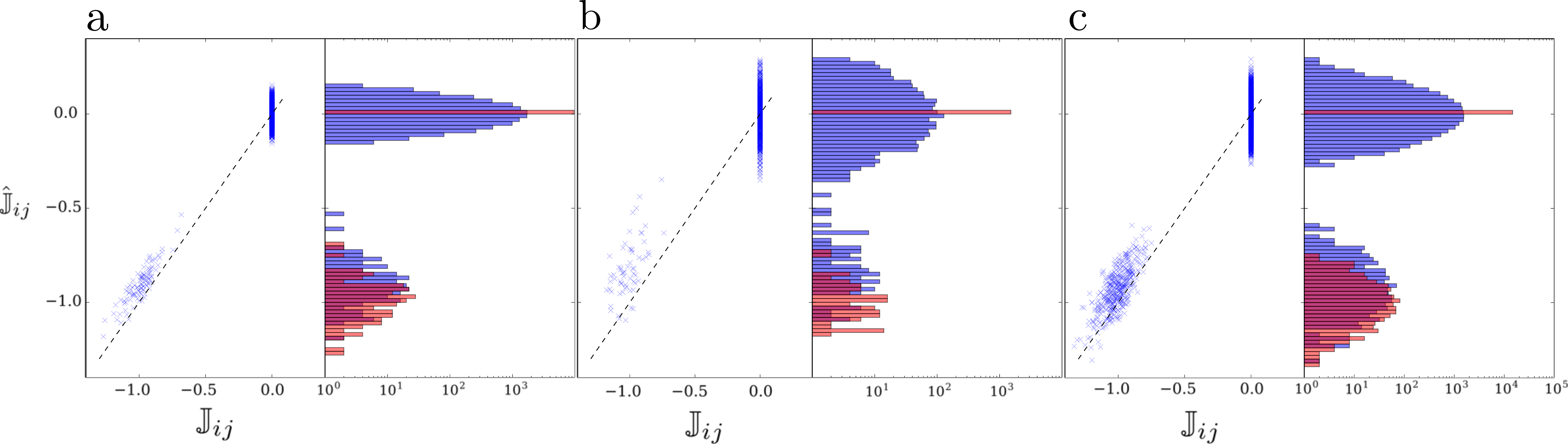}
 \caption{\textbf{Partial inference on large networks.} 
 Partial velocity correlator inference of Eq.~(\ref{eq:est1}), for networks with $n=1000$ agents, $m \approx 100$ of which are accessible to measurement. 
 Blue crosses plot the inferred matrix elements $\hat{\J}_{ij}$ between measurable agents against their real value $\J_{ij}$. Converted into histogram 
 form, the data exhibit a clear separation between low-valued inferences -- corresponding to vanishing couplings -- and higher-valued inferences -- corresponding to existing couplings. Red histograms correspond to the true 
 Jacobian matrix of dynamical flows. The networks are:
 (a) an Erd\H os-R\'enyi network~\cite{Bar16}; (b) a Barab\'asi-Albert network~\cite{Bar16};  (c) a Watts-Strogatz network~\cite{Wat98}. 
 The small but still 
 significant inference imprecision is due to computational limits for generating velocity time series by simulating the dynamics
 of these networks, and not to our inference method. }
 \label{fig:fig4}
\end{figure*}
\subsection{Partial reconstruction from partial measurements.}
Eq.~(\ref{eq:est1}) makes it clear that $\J_{ij}$ can be predicted from 
time series for agents $i$ and $j$ only, and we already mentioned that this enables partial reconstruction of network structures even when 
there are inaccessible agents.
The power of our method 
for partial network structure inference has already been illustrated in Fig.~\ref{fig:fig1} and we next show how it scales to larger systems.

Fig.~\ref{fig:fig4} shows how our inference method correctly differentiates the nonzero network couplings from the vanishing ones, for three 
types of complex networks with $n=1000$ agents, about $m \approx 100$ of which are accessible to measurements. The chosen finite recording time
and the finite noise correlation time result in an uncertainty of the inferred matrix elements $\hat{\J}_{ij}$, however their histogram indicates
a clear separation between low-valued and high-valued $\hat{\J}_{ij}$, i.e. between vanishing and existing direct network couplings. For the existing 
couplings, one furthermore sees that the inferred histogram largely overlaps with the real ones, since
the predicted coupling strengths are quantitatively captured. 

In the Supplementary Information we show a direct comparison of the results of Fig.~\ref{fig:fig4} with predictions from
the method of Ref.~\citen{Ren10} which shows that the latter, applied outside its range of validity to infer the coupling network between 
a subset of measurable agents,
fails in predicting the existing couplings and their coupling strength. Our method thus offers
a significant improvement over existing approaches when not all nodes are accessible to measurement. 


\begin{figure}
 \centering
 \includegraphics[width=0.65\columnwidth]{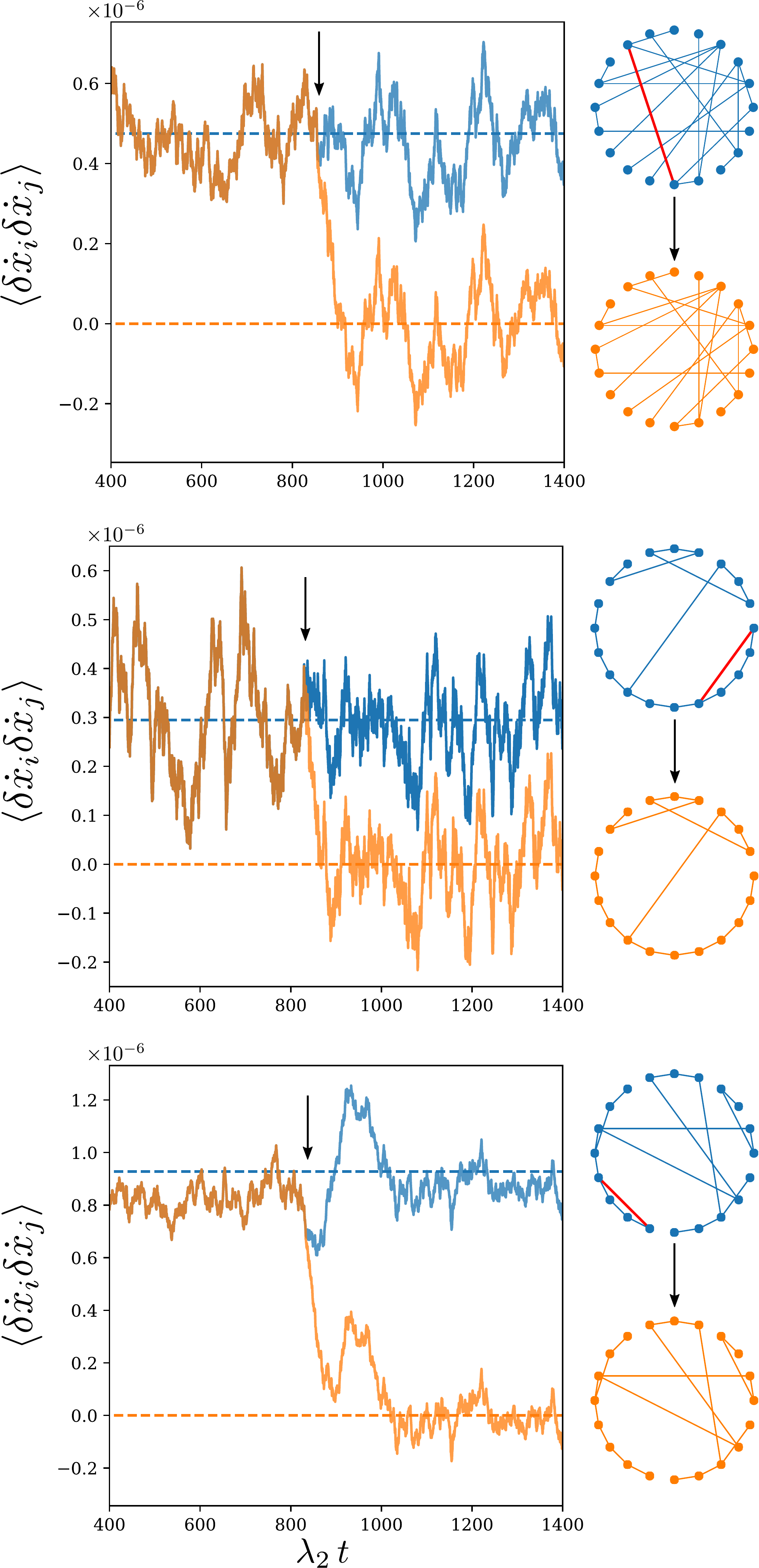}
 \caption{\textbf{Real-time monitoring of dynamically evolving networks.} 
 Time evolution of the agent velocity correlator between agents $i$ and $j$ for three different Watts-Strogatz networks~\cite{Wat98}.
 with $n=20$. The correlator is calculated over sliding time-windows.
 The coupling between $i$
 and $j$ (red edge on the blue networks on the right panels) is removed at time $\lambda_2 t=900$, indicated by the arrow in each panel. Following that 
 topological change, the velocity correlator decreases fast and oscillates around zero (orange curve), well below its behavior
 without the change (blue curve), as predicted by Eqs.~(\ref{eq:dis2}) and (\ref{eq:est1}). In all three cases, the convergence time $\tau_c$ to the 
 new behavior is determined by the smallest nonvanishing Jacobian eigenvalue, $\tau_c \sim \lambda_2^{-1}$, with $\lambda_2=5.84 \times 10^{-3}$, $4 \times 10^{-3}$ and $8.2 \times 10^{-3}$ from top  to bottom.} 
 \label{fig:fig5}
\end{figure}

\subsection{Time-Evolving Networks.} 
We finally show how the noisy agent dynamics directly reflects topological changes in the coupling network structure. From Eqs.~(\ref{eq:dis2}) and (\ref{eq:est1}), disconnecting a network edge between two agents reduces the corresponding velocity correlator, and makes it even 
disappear in the white-noise limit. Recording the noise and calculating the velocity correlator in real-time  enables
to identify topological changes. This is illustrated in Fig.~\ref{fig:fig5} for three Watts-Strogatz networks where one coupling is cut at 
$\lambda_2 t \approx 900$ (see arrows). 
The velocity correlator is quickly reduced compared to the value it has without 
topological change, almost directly reflecting the coupling cut. 
In the Supplementary Information, the calculation of the velocity correlator indicates that transient terms exist, which 
disappear exponentially in time with rates given by the eigenvalues of the network Jacobian matrix $\J$. These terms govern the 
transient behavior following the topological change in Fig.~\ref{fig:fig5}. Therefore,
such topological changes can be identified after a time on the order $t_c \sim \lambda_2^{-1}$ with the smallest nonvanishing
eigenvalue $\lambda_2$ of the Jacobian of dynamical flows after the topological change. 
This reasoning is confirmed in Fig.~\ref{fig:fig5} for three networks with different $\lambda_2$.

\section{Discussion}

We have shown how to infer a coupling network via dynamical monitoring of its agents. The method presents a number of advantages over 
existing approaches. In particular (i) it is nonintrusive; it does not require the ability to inject a perturbation signal locally on specific agents, (ii) 
it allows to reconstruct network structures such as direct couplings and geodesic distances between measurable agents, even when 
only partial measurements are feasible, (iii) it is algorithmically efficient; unlike earlier approaches it does not rely on a matrix inversion as it
directly reconstruct the network matrix; it is therefore easily scalable to larger networks, and (iv) it allows to monitor time-evolving networks
in real-time and in particular to identify when a direct coupling line between two agents is cut. 
The price to pay for these advantages is that velocity, and not position correlators need to be 
measured. One may then think that the resolution in time sampling necessary to extract velocities from
positions would render our method inapplicable in practice. 
In the supplementary material, we show that this is not so, as the time sampling needs only to resolve the time scales $\lambda_\alpha^{-1}$ of the network but not the noise correlation time $\tau_0$. 
This poses only a weak
condition on the resolution one must have to extract velocities from positions. Finally, we also show
in the Supplementary Information that our method is more robust against measurement noise than earlier
ones.
We therefore think that our method will prove to be very beneficial to
infer basic interactions in large networks, which often cannot be fully monitored nor directly probed, or simply to extract partial network
structures, when one does not need to know the full network.  

Classes of network-coupled systems include those with 
higher-dimensional agents with intrinsic, internal dynamics. While not considered in our
numerical illustrations, we believe that our method also applies to such systems in the case of fully symmetrical couplings, i.e. also with respect to the internal degrees of freedom. Network reconstruction in such cases has been attempted based on data-based approaches~\cite{Ero20} and we leave it to future works to illustrate the power of our method in such cases.

As a final remark, when inferring an unknown network from measurement of the dynamics of its agents, 
one may be trying to reconstruct a disconnected network without knowing it. In that case, we show in the Supplementary Information 
that earlier inference methods based on
measurement time series of agents dynamics and their correlators have trouble differentiating between existing and non-existing couplings.
Our method does not suffer from this shortcoming. As such it is able to reconstruct network structures for fully unknown networks, where 
neither all nodes, nor the network connectivity are known {\it a priori}. Future works might apply our method to time series measured from real-world systems.
\\

\newcommand{\hbAppendixPrefix}{A}
\renewcommand{\thefigure}{\hbAppendixPrefix\arabic{figure}}
\setcounter{figure}{0}
\renewcommand{\thetable}{\hbAppendixPrefix\arabic{table}} 
\setcounter{table}{0}
\renewcommand{\theequation}{\hbAppendixPrefix\arabic{equation}} 
\setcounter{equation}{0}
\section*{Appendix A. Methods}

\subsection{Dynamical model.} In our numerical investigations we focus on Eq.~(\ref{eq:dyn0}) with 
\begin{equation}\label{eq:F}
F_i[{\bm x}(t)] = \omega_i - \sum_j a_{ij} \, f(x_i-x_j) .
\end{equation}
Here, $\omega_i\in\mathbb{R}$ are natural frequencies with $\sum_i\omega_i=0$, and 
the interaction between agents is a differentiable function $f\colon\mathbb{R}\to\mathbb{R}$, that is even in its indices $i$ and $j$ and odd in its argument, and $a_{ij} \ge 0$ are unknown elements of the adjacency matrix of the interaction network. 
When the nonvanishing $a_{ij}$ are sufficiently large and numerous, Eq.~(\ref{eq:dyn0}) has at least one 
stable fixed point $\bm{x}^*\in\mathbb{R}^n$, see e.g. Refs.~\citen{Dor13,Del17a}.

For this type of models, the Jacobian matrix of dynamical flows reads
\begin{equation}\label{eq:laplacian}
 \J_{ij}(\bm{x}^*) = \left\{
 \begin{array}{cc}
  -a_{ij}\, \partial_x \, f(x)\big|_{x=x_i^*-x_j^*} \, , & i \ne j \, , \\
  \sum_k a_{ik} \, \partial_x \, f(x)\big|_{x=x_i^*-x_k^*} \, , & i=j \, .
 \end{array}
 \right.
\end{equation}
It is a Laplacian matrix with zero row and column sums of its components, and one vanishing eigenvalue, $\lambda_1=0$, corresponding to 
a constant-component eigenmode, ${\bm u}_1 = (n^{-1/2},n^{-1/2}, \ldots n^{-1/2})$. Eq.~(\ref{eq:laplacian}) makes it clear that $\J({\bm x}^*)$
contains information on both the coupling network and the fixed point $\bm x^*$. 

\subsection{Numerical simulations.} Dynamical series for the agent coordinates and velocities are obtained from 
 Eqs.~(\ref{eq:dyn0}) and (\ref{eq:F}) that are time-evolved following a fourth-order Runge-Kutta algorithm. In most numerical simulations, we
 considered the short correlation time limit, $\lambda_n \tau_0 \ll 1$. Short noise correlation times require even shorter Runge-Kutta 
 time steps for accurate dynamical calculations. This 
 results in rather long computation times for generating velocity time series of sufficient duration, while time series on only one every ten (or more)
 Runge-Kutta times steps are needed as input for our inference method. 
 The generation of these input data requires computation times on the order of a week on twelve Intel\textsuperscript{\textregistered} Xeon\textsuperscript{\textregistered}  Gold 6140 CPU @ 2.30GHz
 for each simulation of the larger networks considered in this article.
Longer computation times would improve the accuracy of our numerical simulations. It is important to understand that 
 this is not a shortcoming of our approach, which infers network structures in a fraction of the time needed to generate its input.
 The convergence of the algorithm to the true Jacobian as longer time series are generated is illustrated in the 
 Supplementary Information. 

\section*{Supplementary Material}
The supplementary Material gives details of the analytical results presented in the main text, considers the case of inference with noise that has long correlation time, shows further numerical illustration of our inference method and also discusses the effect coming from the sampling rates of the time series on the precision of our method.

\section*{Acknowledgments}
This work has been supported by the Swiss National Science Foundation under grants 200020$_-$182050 and P400P2$_-$194359.
RD acknowledges support by ETH Z\"urich funding. \\[-1mm]

\section*{Data Availability}
The data that support the findings of this study are available from the corresponding author upon reasonable request.

\bibliographystyle{unsrt}

\textbf{\large Author contributions}

Numerical investigations and design of result presentation were performed by M.T. All authors contributed to the development of the theory, the
analytical calculations and to editing and writing the manuscript. \\[-1mm]

\textbf{\large Additional information}

\textbf{Supplementary information} accompanies this article. \\[-1mm]

\textbf{Competing financial interests:} The authors declare no competing financial interest.

\end{document}